\documentclass[
aps,
prl,
 amsmath,amssymb,
 reprint
]{revtex4-1}

\usepackage{amsmath}
\numberwithin{figure}{section}
\renewcommand{\thefigure}{\arabic{figure}}

\usepackage{graphicx}
\usepackage{dcolumn}
\usepackage{bm}

\usepackage[utf8]{inputenc}
\usepackage[T1]{fontenc}
\usepackage{mathptmx}
\usepackage{etoolbox}

\makeatletter
\def\@email#1#2{%
 \endgroup
 \patchcmd{\titleblock@produce}
  {\frontmatter@RRAPformat}
  {\frontmatter@RRAPformat{\produce@RRAP{*#1\href{mailto:#2}{#2}}}\frontmatter@RRAPformat}
  {}{}
}%

\usepackage{pdfpages}

\makeatletter
\AtBeginDocument{\let\LS@rot\@undefined}
\makeatother

\makeatother
\begin{document}

\title[]{On the Relation Between the Boson Peak and the Vibrational Phases in Glasses}

\author{Philip Rasmussen}
\author{Søren S. Sørensen$^{*}$}%
 \email{soe@bio.aau.dk}
\affiliation{ 
Department of Chemistry and Bioscience, Aalborg University, Fredrik Bajers Vej 7H, 9220 Aalborg, Denmark
}%

\date{\today}

\begin{abstract}
The boson peak is a vibrational feature commonly found in glassy and disordered systems described as the deviation from the scaling of the Debye model ($g( E)\propto  E ^2 $). Despite much interest, the microscopic nature of the vibrations is only anecdotally discussed. Here we report how the boson peak appears in conjunction with a crossover from localized to extended but non-wave-like and coincide with a maximum in phase coherence and spatial periodicity. Our analyses provide new insights into the microscopic origin of the boson peak across a wide range of glasses.
\end{abstract}

\maketitle


The low energy vibrations in crystals and glasses are markedly different~\cite{Zeller1971ThermalSolids}. This is often observed as an excess of vibrations  in the glassy state beyond that expected by the Debye model (which predicts the density of states to scale as $g( E)\propto  E^2$)~\cite{Schirmacher2024}. This phenomenon is commonly coined as the \textit{boson peak} (BP) and consequently observed as a peak in inelastic scattering from e.g. Raman, X-ray, and neutron spectroscopy~\cite{Shimodaira2005,Kabeya2016-ni}, particularly when plotting the reduced density of states ($g( E) E^{-2}$)~\cite{Parshin1993}. Up to now, the BP has been established as a phenomenon in a large number of different systems of mainly amorphous but very varying chemical origin, e.g. oxides~\cite{Wada2024,Schroeder2004}, chalcogenides~\cite{Novikov1991}, metallic alloys~\cite{Li2008}, polymers~\cite{Zorn2018}, and small organic molecules~\cite{GonzlezJimnez2023,Ruta2010}.

Despite the BP's ubiquity, its microscopic origin remains a longstanding subject of debate. A variety of theoretical frameworks have been proposed, including the soft potential model (SPM)~\cite{Karpov1982,Buchenau1992,Parshin1993}, which attributes the BP to excess quasi-localized vibrational modes (QLVs) associated with anharmonic soft potentials. Other authors argue that heterogenous elasticity~\cite{Schirmacher2006}, that is, structural disorder which gives rise to spatial fluctuations in local elastic moduli, is responsible for the BP. Moreover, interpretations relate the BP to disorder-broadened remnants of crystalline van Hove singularities~\cite{Taraskin2001,Chumakov2015,Hu2023}. Beyond attempts to establish the microscopic origin of the BP, considerable attention has also been devoted to its relation to the Ioffe-Regel crossover. In particular, the BP energy is often found to coincide with the transverse Ioffe-Regel limit~\cite{Beltukov2016,ruffle2008boson,Shintani2008}. Despite the many viewpoints, a complete microscopic understanding of the vibrations responsible of the BP is missing.  A recurring limitation of these studies is that the proposed mechanisms have generally been investigated only in a limited number of model glasses and realistic glass formers, which has raised the question of whether they reflect universal features of glassy solids across a broad variety of glasses, and if yes, what governs the microscopic vibrations related to the BP.

In this \textit{Letter} we therefore investigate the boson peak across an unprecedentedly large range of different glasses, including oxides, chalcogenides, metallic glasses, hybrid organic-inorganic glasses, and glassy water, representing a total of >60 distinct composition-pressure combinations. By adopting such a broad range of glasses, we aim to identify universal features of the vibrational modes around the boson peak, even across systems spanning vastly different chemistries, compositions, atomic densities, and local bonding environments. Through normal mode analysis we assess boson peak energies ($ E_\text{BP}$) as well as the eigenvectors governing the underlying vibrations. Using recent advances in eigenmode analyses we showcase how $ E_\text{BP}$ consistently coincides with a maximum in the acoustic character and spatial periodicity of modes, shedding new light on the nature of the boson peak.

\textit{Methods} - We prepared atomistic models of oxides and modified oxide glasses $x\mathrm{Na_2O}\text{-}(100\text{-}x)\mathrm{SiO_2}$,
$30\mathrm{CaO}\text{-}\,10\mathrm{Al_2O_3}\,\text{-}\,60\mathrm{SiO_2}$,
$x\mathrm{BeO}\text{-}(100\text{-}x)\mathrm{SiO_2}$, chalcogenide glasses ($\text{Ge}_x\text{Se}_{100\text{-}x}$), metallic glasses (Zr-Cu-Al), metal–organic framework glass (based on ZIF-4), and hyperquenched glassy water (HQ-$\mathrm{H_2O})$. Within these, we explore a wide range of compositions and pressure. Detailed description of the glass preparation procedures and compositions are provided in the Supplemental Material~\cite{SI}. We aim to pick simulation protocols that yield realistic models of experimental glasses. To confirm the structural agreement between simulated and experimental glasses, we compute the structure factor and evaluate the $R_{\chi}$-factor as of Wright~\cite{Wright1993} of selected glasses (See Fig.~S1 in the Supplemental Material~\cite{SI}) to find overall good agreement with the majority of selected compositions. We note how glasses which are only achievable within the computer simulation, may still represent meaningful comparison to glasses which are more realistic.

\begin{figure}[!htbp]
    \centering
    \includegraphics[width=0.85\columnwidth]{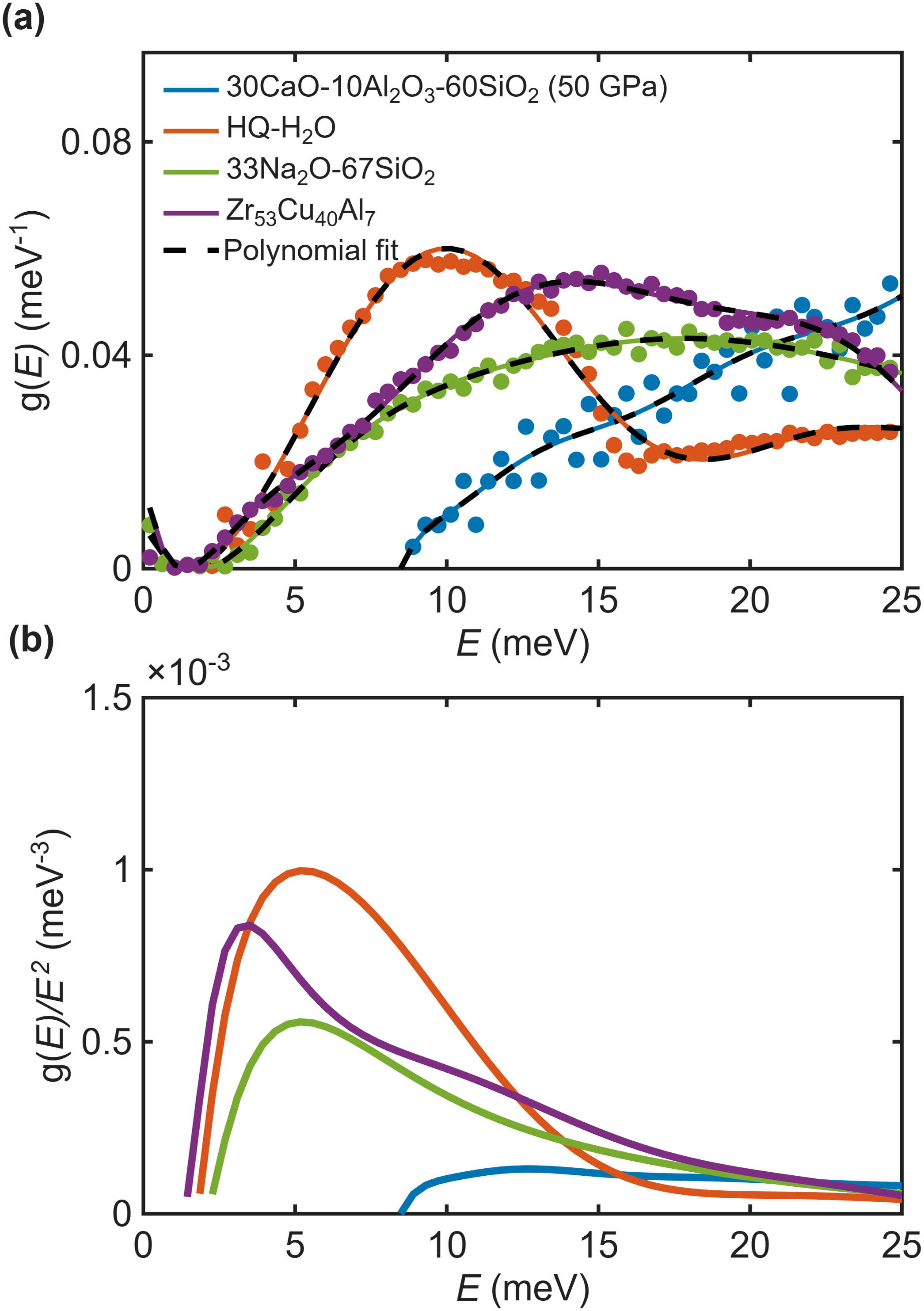}
    \caption{Low-energy vibrational density of states ($g( E)$) for representative computer glasses(a), and the corresponding reduced $g( E)$, $g( E) E^{-2}$, revealing the boson peak (b). The $g( E)$ was obtained by binning the eigenenergies and averaging over six independent realizations for each composition. Dashed lines in (a) represent 10th-order polynomial fits to the averaged $g( E)$, which were subsequently used to compute the $g( E) E^{-2}$ shown in (b). Pronounced variations in both the spectral shape and the distribution of low-energy modes are observed across systems, highlighting the strong influence of chemistry and local bonding environments on the vibrational spectrum.}
    \label{vdos_reddos}
\end{figure}

\textit{Results and discussion} - The great diversity in chemistry across investigated glasses yield markedly different BP signatures with substantial variation in intensity, width, and position. To identify the BP energy, we calculate the vibrational density of states ($g( E)$) from normal-mode analysis (see Methods in Supplemental Material~\cite{SI} for details). We then reveal the BP from the reduced $g( E)$ as $g( E) E^{-2}$. In Fig.~\ref{vdos_reddos}(a), we highlight the wide variation in the low-energy vibrational spectra across several glass families. When considering the reduced density of states, $g( E) E^{-2}$ (Fig.~\ref{vdos_reddos}(b)), these differences become even more pronounced, with significant variations in peak position, intensity and overall shape of the BP. These observations highlight the strong influence of microscopic structure and bonding on the vibrational response of glasses.

Despite this apparent diversity, the common occurrence of the boson peak across all systems suggests an underlying universal mechanism. This raises a central question: which features do vibrations at, or in vicinity of, the BP share? To address this, we first examine the connection between the BP energy and the elastic properties of the investigated glasses.

\begin{figure}[!t]
    \centering
    \includegraphics[width=\columnwidth]{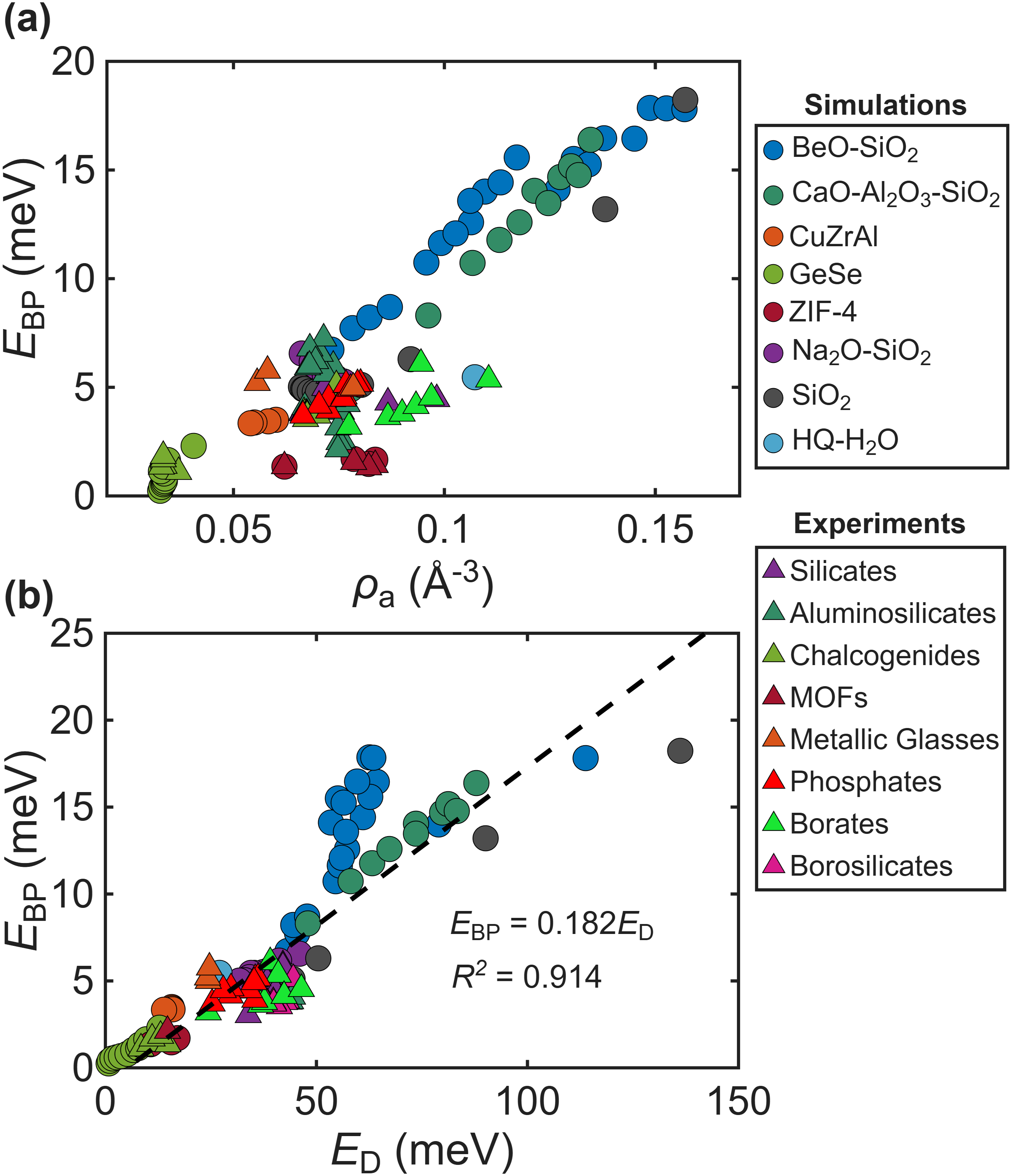}
    \caption{Scaling of the boson peak energy ($ E_\mathrm{BP}$) with (a) atomic number density ($\rho_\mathrm{A}$)  and (b) Debye energy ($ E_\mathrm{D}$) for simulated glasses and selected experimental data. The dashed line is linear fit to the data points, simulations and experimental glasses included, yielding $ E_\mathrm{BP}\approx0.182\, E_\mathrm{D}$ ($R^2=0.914$), highlighting the close connection between the boson peak and the underlying elastic properties of disordered solids. Experimental measurements are obtained from~\cite{aSawamura2018,bndo2018,cdeMacedo2018,dLimbach2015,eSawamura2018,fPan2021,g1983,hMncke2016,refid23Srensen2019,jGriebenow2018,kDa2010,lLe2017,mWang2012,nStepniewska2020,oRamos2004,pScarponi2004,qOta1978,rZhao2012}.}
    \label{bp_debye}
\end{figure}

In Fig.~\ref{bp_debye}(a) we compile both simulated and experimental data from the literature to reveal an overall strong correlation between $ E_\text{BP}$ and atomic number density ($\rho_\text{a}$), with denser glasses exhibiting higher $ E_\text{BP}$. This suggests that $ E_\text{BP}$ is closely linked to the underlying packing and connectivity of the glass network. The observed trend is consistent with recent theoretical work proposing a common origin of the BP in both crystalline and amorphous solids, in which the $ E_\text{BP}$ increases with increasing atomic packing density~\cite{Baggioli2019}

To now assess the combined role of elasticity and atomic density we compute the Debye energy ($ E_{\text{D}}$),

\begin{equation}\label{debye_frequency}
 E_\text{D}^3 = 6\pi^2 \hbar^3 \rho_a v_\text{D}^3,
\end{equation}

where $\rho_\text{a}$ is the atomic number density, $\hbar$ is the reduced Planck constant, and the Debye sound velocity is given by $v_\text{D}^{-3}=((1/v_\text{L}^3)+(2/v_\text{T}^3))/3$. As shown in Fig.~\ref{bp_debye}(b), both simulated and experimental glasses exhibit a strong correlation between $ E_\text{BP}$ and $ E_\text{D}$, yielding approximately $ E_\text{BP}\simeq0.18\, E_\text{D}$, in good agreement with previous studies (reporting $ E_\text{BP}\simeq0.1 - 0.2\, E_\text{D}$~\cite{ruffle2008boson,monaco2009anomalous,baldi2009connection}). This observation suggests that a common elastic energy scale underlies the BP across chemically and structurally diverse glass families.

While scaling demonstrates that elastic properties can largely predict $ E_\mathrm{BP}$, it does not reveal the microscopic signature of the underlying vibrational modes. In the Debye model, vibrations are assumed to be fully acoustic plane waves. This is well-known to be a flawed description of real solids~\cite{Lerner2016}. However, only recently have metrics  been developed to assess the vibrational periodicity of disordered materials, e.g. to provide a quantification of the degree of local phase coherence and periodicity  of the microscopic vibrations~\cite{Bell1975,Seyf2016}. Nonetheless, such analyses have been limited to few systems and have yet to see wider adoption~\cite{Seyf2016,Seyf2017,Huang2021}.

\begin{figure*}[!htbp]
    \centering
    \includegraphics[width=1\textwidth]{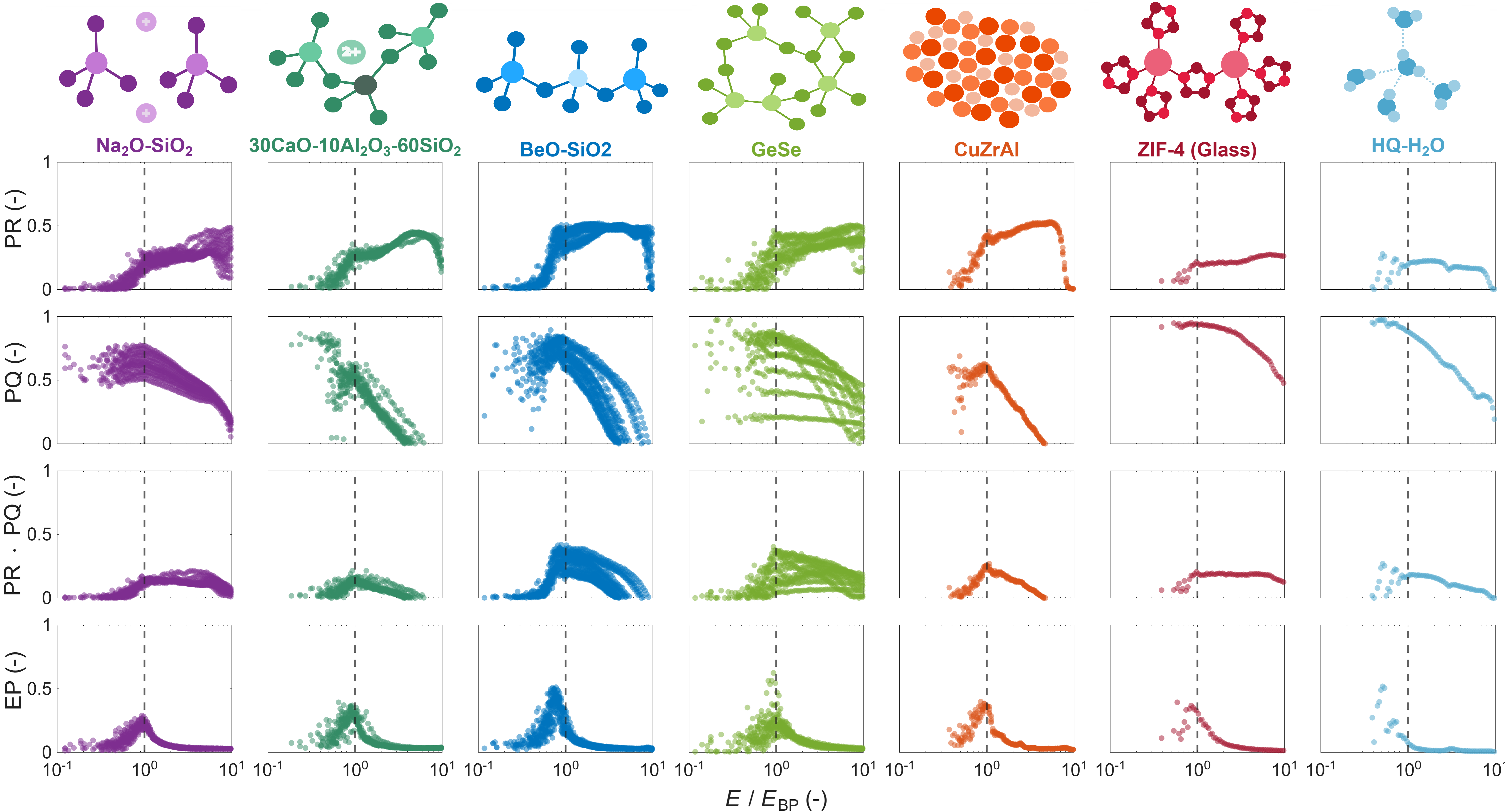}
    \caption{Modal characteristics rescaled by $ E/ E_\text{BP}$ across five different glass families and 60 compositions, including modified oxides, chalcogenides, metallic glasses, hybrid organic-inorganic (MOF) glasses, and glassy water, spanning a wide range of bonding environments including covalent networks, modifier-induced ionic-covalent interactions, metallic bonding, coordination networks, and hydrogen bonding. Representative schematics of local bonding motifs are shown above, highlighting the vast differences of chemistry and bonding characteristics across glass families. A common characteristic feature is observed at $\sim E_\text{BP}$ for (first row) participation ratio (PR), (second row) local phase coherence (PQ), (third row) the combined descriptor PR$\cdot$PQ, and (fourth row) spatial periodicity (EP). The vertical dashed line indicates $ E= E_\text{BP}$. PR increases towards $ E_\text{BP}$, indicating increasingly extended vibrational motion. PQ and EP exhibit pronounced maxima near $ E_\mathrm{BP}$, followed by a rapid decrease, indicating that the boson peak coincides with the most coherent and wave-like vibrations sustained by the disordered structure.}\label{PRPQEP}
\end{figure*}

In the present work we utilize a number of metrics to broadly probe the vibrations across the studied glasses. First, to study localization of atomic vibrations we calculate the \textit{participation ratio} (PR) for each mode as,

\begin{equation}\label{PR}
\mathrm{PR}_n = \left( N \sum_{i=1}^{N} \left| \vec{\mathbf{e}}_{i,n} \right|^4 \right)^{-1},
\end{equation}

where $\mathbf{e}_{i,n}$ is the eigenvector associated with atom $i$ of the $n$-th mode and $N$ is the number of atoms. The PR measures the fraction of atoms that contribute to a vibrational mode, ranging from $N^{-1}$ for a fully localized vibration to unity for fully extended vibrations. In Fig.~\ref{PRPQEP} (top row) we show the PR of each glass family as a function of the reduced energy $ E/ E_\text{BP}$. We find modes at the lowest studied energies to be strongly localized confirming previous suggestions on how this energy region hosts quasilocalized vibrational modes~\cite{Shimada2020}. Beyond their localized character, quasilocalized vibrations are distinguished by their contribution to the low-energy nonphononic spectrum, which exhibits the universal density of states scaling $g( E)\propto E^4$~\cite{Lerner2016,Richard2020}. As the energy of the modes increases, a distinct crossover appears at approximately $ E_\text{BP}$ for all investigated glasses, beyond which the participation ratio stabilizes. The change at this particular energy (i.e.
$ E_\text{BP}$) suggests that the BP corresponds to a characteristic transition in the spatial extent of vibrational excitations in glasses.

While the PR quantifies the spatial extent of vibrations, it does not provide information about the direction or coherence of atomic motion. Since the Debye description of solids is based on long-wavelength acoustic waves characterized by coherent neighboring displacements, deviations from Debye behavior are expected to be accompanied by changes in local vibrational coherence. To study this, we calculate the so-called \textit{phase quotient} (PQ) introduced by Bell and Hibbins-Butler~\cite{Bell1975},

\begin{equation}\label{PQ}
       \mathrm{PQ}_n = \frac{\sum_{m} \vec{\mathbf{e}}_{i,n}\cdot \vec{\mathbf{e}}_{j,n}}{\sum_{m}|\vec{\mathbf{e}}_{i,n}\cdot \vec{\mathbf{e}}_{j,n}|},
\end{equation}

where the summation is performed over all $m$ nearest-neighbor pairs $(i,j)$ in the system and $\mathbf e_{i,n}$ and $\mathbf e_{j,n}$ denote the corresponding eigenvector components. This provides a spatial average of local phase coherence in vibrational modes, ranging from 1 (in-phase, acoustic-like) to -1 (out of phase, optical-like). In Fig.~\ref{PRPQEP} (second row) we plot the PQ of all glasses as a function of reduced energy $ E/ E_\mathrm{BP}$. We find that low-energy vibrations exhibit substantial local phase coherence. As energy increases, coherence increases and, like PR, reaches a pronounced maximum in the vicinity of $ E_\text{BP}$. Beyond this point, the PQ decreases rapidly, indicating a progressive loss of local phase coherence. The occurrence of this crossover at approximately $ E_\text{BP}$ across all investigated glasses suggests a close connection between the boson peak and loss of vibrational coherence.

To capture both the spatial extent and local phase coherence of modes simultaneously, we consider the product $\mathrm{PR}_n\cdot \mathrm{PQ}_n$. As shown in Fig.~\ref{PRPQEP} (third row), this quantity exhibits a broad maximum or enhanced plateau in the vicinity of $ E_\text{BP}$ for most glasses, indicating that vibrational modes near the BP are characterized by a combination of relatively large spatial extent and strong local coherence. However, this combined value only describes coherence within a nearest-neighbor perspective as neither PQ nor PR directly probe whether the vibrations follow an underlying wave-like pattern. 

To examine the extent to which vibrational modes follow an underlying periodic wave-like pattern, we employ the recently developed metric of \textit{eigenvector periodicity} (EP)~\cite{Seyf2016}. EP quantifies how closely a vibrational eigenmode can be matched to an optimally chosen periodic reference mode. Specifically,

\begin{equation} 
\mathrm{EP}_n = \frac{\Psi_n(\vec{\mathbf{k}}^{\prime},\phi^{\prime})} {\Psi_{n\mathrm{,ref}}(\vec{\mathbf{k}}^{\prime},\phi^{\prime})}, 
\end{equation}

where $\Psi_n$ is a measure of the degree of periodicity of mode $n$, and $\Psi_{n,\mathrm{ref}}$ is the corresponding reference mode with perfectly periodic character. The quantities $\vec{\mathbf{k}}^{\prime}$ and $\phi^{\prime}$ denote the wave vector and phase that maximize the periodicity measure $\Psi(\vec{\mathbf{k}},\phi)$. A detailed derivation is provided in Ref.~\cite{Seyf2016} and summarized in the Supplemental Material~\cite{SI}. By construction, $\mathrm{EP}=0$ corresponds to the absence of periodic wave-like order, whereas $\mathrm{EP}=1$ denotes a perfectly periodic wave-like mode, e.g. a fully acoustic mode.

The resulting EP spectra are shown in Fig.~\ref{PRPQEP} (fourth row). At low energies, all investigated glasses exhibit relatively small EP values, indicating that modes in the $ E\rightarrow0$ limit are governed by vibrations that cannot be described as well-defined periodic waves despite their relatively high local phase coherence (as evidenced by their positive PQ values). As mode energy increases, EP rises steadily and reaches a pronounced maximum near $ E_\text{BP}$, followed by a gradual decrease of EP. This highlights a strong connection between the BP and periodic wave-like order of its underlying vibrational phase. Furthermore, the occurrence of the EP maximum near $ E_\mathrm{BP}$ across all glass families suggests that the boson peak represents a characteristic vibrational regime in which wave-like order is maximized.

Taken together, the PR, PQ, and EP analyses reveal a consistent picture of the vibrational behavior of low energy excitations in glasses. While PR identifies a crossover in the spatial extent of vibrational excitations, PQ and EP exhibit pronounced maxima near $ E_\text{BP}$. These observations indicate that the boson peak corresponds to a characteristic vibrational regime in which modes simultaneously exhibit substantial spatial extent, maximal local phase coherence as well as maximal wave-like order. Furthermore, the combined observations suggest that the low-energy excitations in the lower $ E$ limit are spatially localized, locally coherent, yet lack a well-defined periodic wave-like structure. Their characteristic core sizes are consistent with previously reported QLVs in structural glasses~\cite{Richard2020,Lerner2016,Moriel2024,Gartner2016,Kapteijns2020} (see Fig.~S2(a) in the Supplemental Material~\cite{SI}).

The emergence of a common crossover energy in the PR, PQ, and EP analyses suggests that the BP is linked to a fundamental change in the nature of vibrations in glasses near $ E_{\text{BP}}$. This raises the question of whether the BP can be connected to an underlying acoustic energy scaling. Richard \emph{et al.}~\cite{Richard2020} reported a PR crossover analogous to that observed here, occurring at the first shear phonon band energy ($ E_{v_\text{T}}=h v_\text{T}/L$). Expanding on this idea, we rescale the PR, PQ, and EP by the first shear and longitudinal phonon band energies, $ E_{v_\text{T}}$ and $ E_{v_\text{L}}$, where $v_\text{T}$ and $v_\text{L}$ are the shear and longitudinal sound velocities, respectively, $L$ is the simulation-box length, and $h$ is the Planck constant. Remarkably, the characteristic PR, PQ and EP crossovers collapse near $ E/ E_{v_\text{T}}\approx1$ (See Fig.~S3 in the Supplemental Material~\cite{SI}), whereas no analogous correlation is observed for $ E_{v_\text{L}}$ (See Fig.~S4 in the Supplemental Material~\cite{SI}). This highlights a connection between shear waves and the BP, an observation consistent with the frequently proposed connection between the BP and the transverse Ioffe-Regel crossover~\cite{Shintani2008,Ren2021, Hu2022,Xi2026,Park2026}, the point at which transverse acoustic excitations become so strongly scattered that the concept of a well-defined propagating shear wave breaks down.

In summary, the present work simulated a large number of realistic glasses with great diversity of chemistry and local bonding environments to study the universality of vibrations around the boson peak. By combining analyses of the $g( E)$ with a detailed analysis of eigenmode characteristics, we identified a common underlying behavior that is not evident from the $g( E)$ spectra alone. 
Specifically, we find that the boson peak energy ($ E_\text{BP}$) across the glass families marks a crossover at which both the phase coherence and spatial periodicity attain pronounced maxima, while the participation ratio increases from low values associated with localized vibrations and saturates near $ E_\text{BP}$. We attribute this behavior to a transition from localized vibrational modes to an extended but non-wave-like vibrational regime.
Above $ E_\text{BP}$, modes remain spatially extended but progressively lose their acoustic, plane-wave-like character, demonstrating that breakdown of phonon-like vibrations is governed by disorder-induced loss of coherence rather than being driven by localization. This behavior is consistent with earlier studies linking the boson peak to the transverse Ioffe-Regel energy.
Remarkably, the observed crossover is found in all studied systems despite vast differences in chemistry and local bonding environments, supporting a common universality in the origin of the boson peak in disordered solids.

\section*{Author Declarations}
\subsection*{Conflicts of interest}
The authors declare that they have no competing interests.

\section*{Acknowledgments}
The authors acknowledge computational resources provided by CLAAUDIA and the BioCloud HPC at Aalborg University. We acknowledge Han Liu (Sichuan University) and Mathieu Bauchy (University of California, Los Angeles) for sharing and assistance in setting up the pair potential used for the GeSe simulations~\cite{Liu2026}.


\section*{Data Availability Statement}

The data that support the findings of this study are available from
the corresponding author upon reasonable request.


\section*{REFERENCES}

\bibliography{aipsamp}

@article{Novikov1991,
  title = {A correlation between low-energy vibrational spectra and first sharp diffraction peak in chalcogenide glasses},
  volume = {77},
  ISSN = {0038-1098},
  url = {http://dx.doi.org/10.1016/0038-1098(91)90341-R},
  DOI = {10.1016/0038-1098(91)90341-r},
  number = {3},
  journal = {Solid State Communications},
  publisher = {Elsevier BV},
  author = {Novikov,  V.N. and Sokolov,  A.P.},
  year = {1991},
  month = Jan,
  pages = {243–247}
}

@article{Li2008,
  title = {Study on the boson peak in bulk metallic glasses},
  volume = {104},
  ISSN = {1089-7550},
  url = {http://dx.doi.org/10.1063/1.2948926},
  DOI = {10.1063/1.2948926},
  number = {1},
  journal = {Journal of Applied Physics},
  publisher = {AIP Publishing},
  author = {Li,  Yong and Yu,  Peng and Bai,  H. Y.},
  year = {2008},
  pages={013520}
  
}

@article{Wada2024,
  title = {Absorption dispersion below boson peak frequency in oxide glasses studied by THz-time domain spectroscopy},
  volume = {135},
  ISSN = {1089-7550},
  url = {http://dx.doi.org/10.1063/5.0191384},
  DOI = {10.1063/5.0191384},
  number = {8},
  journal = {Journal of Applied Physics},
  pages={085108},
  publisher = {AIP Publishing},
  author = {Wada,  Osamu and Ramachari,  Doddoji and Yang,  Chan-Shan and Uchino,  Takashi and Pan,  Ci-Ling},
  year = {2024},
  month = Feb 
}

@article{Schroeder2004,
  title = {Raman scattering and Boson peaks in glasses: temperature and pressure effects},
  volume = {349},
  ISSN = {0022-3093},
  url = {http://dx.doi.org/10.1016/j.jnoncrysol.2004.08.265},
  DOI = {10.1016/j.jnoncrysol.2004.08.265},
  journal = {Journal of Non-Crystalline Solids},
  publisher = {Elsevier BV},
  author = {Schroeder,  John and Wu,  Weimin and Apkarian,  Jacob L. and Lee,  Mierie and Hwa,  Luu-Gen and Moynihan,  Cornelius T.},
  year = {2004},
  month = Dec,
  pages = {88–97}
}

@article{Zorn2018,
  title = {Anomalies in the low frequency vibrational density of states for a polymer with intrinsic microporosity – the Boson peak of PIM-1},
  volume = {20},
  ISSN = {1463-9084},
  url = {http://dx.doi.org/10.1039/c7cp07141h},
  DOI = {10.1039/c7cp07141h},
  number = {3},
  journal = {Physical Chemistry Chemical Physics},
  publisher = {Royal Society of Chemistry (RSC)},
  author = {Zorn,  Reiner and Yin,  Huajie and Lohstroh,  Wiebke and Harrison,  Wayne and Budd,  Peter M. and Pauw,  Brian R. and B\"{o}hning,  Martin and Sch\"{o}nhals,  Andreas},
  year = {2018},
  pages = {1355–1363}
}

@article{GonzlezJimnez2023,
  title = {Understanding the emergence of the boson peak in molecular glasses},
  volume = {14},
  ISSN = {2041-1723},
  url = {http://dx.doi.org/10.1038/s41467-023-35878-6},
  DOI = {10.1038/s41467-023-35878-6},
  number = {1},
  journal = {Nature Communications},
  publisher = {Springer Science and Business Media LLC},
  author = {González-Jiménez,  Mario and Barnard,  Trent and Russell,  Ben A. and Tukachev,  Nikita V. and Javornik,  Uroš and Hayes,  Laure-Anne and Farrell,  Andrew J. and Guinane,  Sarah and Senn,  Hans M. and Smith,  Andrew J. and Wilding,  Martin and Mali,  Gregor and Nakano,  Motohiro and Miyazaki,  Yuji and McMillan,  Paul and Sosso,  Gabriele C. and Wynne,  Klaas},
  year = {2023},
  month = Jan,
  pages = {215}
}

@article{Ruta2010,
  title = {Communication: High-frequency acoustic excitations and boson peak in glasses: A study of their temperature dependence},
  volume = {133},
  ISSN = {1089-7690},
  url = {http://dx.doi.org/10.1063/1.3460815},
  DOI = {10.1063/1.3460815},
  number = {4},
  pages = {041101},
  journal = {The Journal of Chemical Physics},
  publisher = {AIP Publishing},
  author = {Ruta,  B. and Baldi,  G. and Giordano,  V. M. and Orsingher,  L. and Rols,  S. and Scarponi,  F. and Monaco,  G.},
  year = {2010},
  
}

@inbook{Schirmacher2024,
  title = {Vibrational excitations in disordered solids},
  ISBN = {9780323914086},
  url = {http://dx.doi.org/10.1016/B978-0-323-90800-9.00166-9},
  DOI = {10.1016/b978-0-323-90800-9.00166-9},
  booktitle = {Encyclopedia of Condensed Matter Physics},
  publisher = {Elsevier},
  author = {Schirmacher,  Walter and Ruocco,  Giancarlo},
  year = {2024},
  pages = {298–317}
}

@article{Shimodaira2005,
  title = {Effects of fictive temperature and halogen doping on the boson peak in silica glass},
  volume = {71},
  ISSN = {1550-235X},
  url = {http://dx.doi.org/10.1103/PhysRevB.71.024209},
  DOI = {10.1103/physrevb.71.024209},
  number = {2},
  journal = {Physical Review B},
  publisher = {American Physical Society (APS)},
  pages={024209},
  author = {Shimodaira,  N. and Saito,  K. and Hiramitsu,  N. and Matsushita,  S. and Ikushima,  A. J.},
  year = {2005},
  month = Jan 
}

@ARTICLE{Kabeya2016-ni,
  title     = "Boson peak dynamics of glassy glucose studied by integrated
               terahertz-band spectroscopy",
  author    = "Kabeya, Mikitoshi and Mori, Tatsuya and Fujii, Yasuhiro and
               Koreeda, Akitoshi and Lee, Byoung Wan and Ko, Jae-Hyeon and
               Kojima, Seiji",
  journal   = "Phys. Rev. B.",
  publisher = "American Physical Society (APS)",
  pages={224204},
  volume    =  94,
  number    =  22,
  month     =  dec,
  year      =  {2016},
}

@article{Karpov1982,
  title = {Atomic tunneling states and low-temperature anomalies of thermal properties in amorphous materials},
  volume = {44},
  ISSN = {0038-1098},
  url = {http://dx.doi.org/10.1016/0038-1098(82)90866-3},
  DOI = {10.1016/0038-1098(82)90866-3},
  number = {3},
  journal = {Solid State Communications},
  publisher = {Elsevier BV},
  author = {Karpov,  V.G. and Klinger,  M.I. and Ignatiev,  P.N.},
  year = {1982},
  month = Oct,
  pages = {333–337}
}

@article{Parshin1993,
  title = {Soft potential model and universal properties of glasses},
  volume = {T49A},
  ISSN = {1402-4896},
  url = {http://dx.doi.org/10.1088/0031-8949/1993/T49A/030},
  DOI = {10.1088/0031-8949/1993/t49a/030},
  journal = {Physica Scripta},
  publisher = {IOP Publishing},
  author = {Parshin,  D A},
  year = {1993},
  month = Jan,
  pages = {180–185}
}

@article{Buchenau1992,
  title = {Interaction of soft modes and sound waves in glasses},
  volume = {46},
  ISSN = {1095-3795},
  url = {http://dx.doi.org/10.1103/PhysRevB.46.2798},
  DOI = {10.1103/physrevb.46.2798},
  number = {5},
  journal = {Physical Review B},
  publisher = {American Physical Society (APS)},
  author = {Buchenau,  U. and Galperin,  Yu. M. and Gurevich,  V. L. and Parshin,  D. A. and Ramos,  M. A. and Schober,  H. R.},
  year = {1992},
  month = Aug,
  pages = {2798–2808}
}

@article{Schirmacher2006,
  title = {Thermal conductivity of glassy materials and the “boson peak”},
  volume = {73},
  ISSN = {1286-4854},
  url = {http://dx.doi.org/10.1209/epl/i2005-10471-9},
  DOI = {10.1209/epl/i2005-10471-9},
  number = {6},
  journal = {Europhysics Letters (EPL)},
  publisher = {IOP Publishing},
  author = {Schirmacher,  W},
  year = {2006},
  month = Mar,
  pages = {892–898}
}

@article{Taraskin2001,
  title = {Origin of the Boson Peak in Systems with Lattice Disorder},
  volume = {86},
  ISSN = {1079-7114},
  url = {http://dx.doi.org/10.1103/PhysRevLett.86.1255},
  DOI = {10.1103/physrevlett.86.1255},
  number = {7},
  journal = {Physical Review Letters},
  publisher = {American Physical Society (APS)},
  author = {Taraskin,  S. N. and Loh,  Y. L. and Natarajan,  G. and Elliott,  S. R.},
  year = {2001},
  month = Feb,
  pages = {1255–1258}
}

@article{Chumakov2015,
  title = {Relation between the boson peak in glasses and van Hove singularity in crystals},
  volume = {96},
  ISSN = {1478-6443},
  url = {http://dx.doi.org/10.1080/14786435.2015.1108528},
  DOI = {10.1080/14786435.2015.1108528},
  number = {7-9},
  journal = {Philosophical Magazine},
  publisher = {Informa UK Limited},
  author = {Chumakov,  Aleksandr I. and Monaco,  Giulio and Han,  Xuemeng and Xi,  Li and Bosak,  Alexey and Paolasini,  Luigi and Chernyshov,  Dmitry and Dyadkin,  Vadim},
  year = {2015},
  month = Nov,
  pages = {743–753}
}

@article{Beltukov2016,
  title = {Boson peak and Ioffe-Regel criterion in amorphous siliconlike materials: The effect of bond directionality},
  volume = {93},
  ISSN = {2470-0053},
  url = {http://dx.doi.org/10.1103/PhysRevE.93.023006},
  DOI = {10.1103/physreve.93.023006},
  number = {2},
  journal = {Physical Review E},
  publisher = {American Physical Society (APS)},
  author = {Beltukov,  Y. M. and Fusco,  C. and Parshin,  D. A. and Tanguy,  A.},
  year = {2016},
  pages={023006},
  month = Feb 
}

@article{Shintani2008,
  title = {Universal link between the boson peak and transverse phonons in glass},
  volume = {7},
  ISSN = {1476-4660},
  url = {http://dx.doi.org/10.1038/nmat2293},
  DOI = {10.1038/nmat2293},
  number = {11},
  journal = {Nature Materials},
  publisher = {Springer Science and Business Media LLC},
  author = {Shintani,  Hiroshi and Tanaka,  Hajime},
  year = {2008},
  month = Oct,
  pages = {870–877}
}

@article{Hu2023,
  title = {Universality of stringlet excitations as the origin of the boson peak of glasses with isotropic interactions},
  volume = {5},
  ISSN = {2643-1564},
  url = {http://dx.doi.org/10.1103/PhysRevResearch.5.023055},
  DOI = {10.1103/physrevresearch.5.023055},
  number = {2},
  journal = {Physical Review Research},
  publisher = {American Physical Society (APS)},
  author = {Hu,  Yuan-Chao and Tanaka,  Hajime},
  year = {2023},
  pages={023055},
  month = Apr 
}

@article{Bell1975,
  title = {Acoustic and optical modes in vitreous silica,  germania and beryllium fluoride},
  volume = {8},
  ISSN = {0022-3719},
  url = {http://dx.doi.org/10.1088/0022-3719/8/6/009},
  DOI = {10.1088/0022-3719/8/6/009},
  number = {6},
  journal = {Journal of Physics C: Solid State Physics},
  publisher = {IOP Publishing},
  author = {Bell,  R J and Hibbins-Butler,  D C},
  year = {1975},
  month = Mar,
  pages = {787–792}
}

@misc{Liu2026,
  author = {Liu, H. and Bauchy, M.},
  title = {Machine Learning-Aided Development of Empirical Force-Fields for Chalcogenide Glasses.},
  year = {2026},
  publisher = {GitHub},
  journal = {GitHub repository},
  howpublished = {\url{https://github.com/SOFT-AI-Lab/GeSeForcefield}}
}

@article{Richard2020,
  title = {Universality of the Nonphononic Vibrational Spectrum across Different Classes of Computer Glasses},
  volume = {125},
  ISSN = {1079-7114},
  url = {http://dx.doi.org/10.1103/PhysRevLett.125.085502},
  DOI = {10.1103/physrevlett.125.085502},
  number = {8},
  journal = {Physical Review Letters},
  publisher = {American Physical Society (APS)},
  author = {Richard,  David and González-López,  Karina and Kapteijns,  Geert and Pater,  Robert and Vaknin,  Talya and Bouchbinder,  Eran and Lerner,  Edan},
  year = {2020},
  pages = {085502},
  month = Aug 
}

@article{Moriel2024,
  title = {Boson peak in the vibrational spectra of glasses},
  volume = {6},
  ISSN = {2643-1564},
  url = {http://dx.doi.org/10.1103/PhysRevResearch.6.023053},
  DOI = {10.1103/physrevresearch.6.023053},
  number = {2},
  journal = {Physical Review Research},
  publisher = {American Physical Society (APS)},
  author = {Moriel,  Avraham and Lerner,  Edan and Bouchbinder,  Eran},
  year = {2024},
  pages = {023053},
  month = Apr 
}

@article{Lerner2016,
  title = {Statistics and Properties of Low-Frequency Vibrational Modes in Structural Glasses},
  volume = {117},
  ISSN = {1079-7114},
  url = {http://dx.doi.org/10.1103/PhysRevLett.117.035501},
  DOI = {10.1103/physrevlett.117.035501},
  number = {3},
  journal = {Physical Review Letters},
  publisher = {American Physical Society (APS)},
  author = {Lerner,  Edan and D\"{u}ring,  Gustavo and Bouchbinder,  Eran},
  year = {2016},
  pages={099901}
  
}

@article{Seyf2016,
  title = {A method for distinguishing between propagons,  diffusions,  and locons},
  volume = {120},
  ISSN = {1089-7550},
  url = {http://dx.doi.org/10.1063/1.4955420},
  DOI = {10.1063/1.4955420},
  number = {2},
  journal = {Journal of Applied Physics},
  publisher = {AIP Publishing},
  author = {Seyf,  Hamid Reza and Henry,  Asegun},
  year = {2016},
  pages = {025101}
  
}

@article{Gartner2016,
  title = {Nonlinear modes disentangle glassy and Goldstone modes in structural  glasses},
  volume = {1},
  ISSN = {2542-4653},
  url = {http://dx.doi.org/10.21468/SciPostPhys.1.2.016},
  DOI = {10.21468/scipostphys.1.2.016},
  number = {2},
  pages = {016},
  journal = {SciPost Physics},
  publisher = {Stichting SciPost},
  author = {Gartner,  Luka and Lerner,  Edan},
  year = {2016},
  month = Dec 
}

@article{Kapteijns2020,
  title = {Nonlinear quasilocalized excitations in glasses: True representatives of soft spots},
  volume = {101},
  ISSN = {2470-0053},
  pages = {032130},
  url = {http://dx.doi.org/10.1103/PhysRevE.101.032130},
  DOI = {10.1103/physreve.101.032130},
  number = {3},
  journal = {Physical Review E},
  publisher = {American Physical Society (APS)},
  author = {Kapteijns,  Geert and Richard,  David and Lerner,  Edan},
  year = {2020},
  month = Mar 
}

@article{Ren2021,
  title = {Boson-peak-like anomaly caused by transverse phonon softening in strain glass},
  volume = {12},
  ISSN = {2041-1723},
  pages = {5755},
  url = {http://dx.doi.org/10.1038/s41467-021-26029-w},
  DOI = {10.1038/s41467-021-26029-w},
  number = {1},
  journal = {Nature Communications},
  publisher = {Springer Science and Business Media LLC},
  author = {Ren,  Shuai and Zong,  Hong-Xiang and Tao,  Xue-Fei and Sun,  Yong-Hao and Sun,  Bao-An and Xue,  De-Zhen and Ding,  Xiang-Dong and Wang,  Wei-Hua},
  year = {2021},
  month = Oct 
}

@article{Hu2022,
  title = {Origin of the boson peak in amorphous solids},
  volume = {18},
  ISSN = {1745-2481},
  url = {http://dx.doi.org/10.1038/s41567-022-01628-6},
  DOI = {10.1038/s41567-022-01628-6},
  number = {6},
  journal = {Nature Physics},
  publisher = {Springer Science and Business Media LLC},
  author = {Hu,  Yuan-Chao and Tanaka,  Hajime},
  year = {2022},
  
  pages = {669–677}
}

@article{Xi2026,
  title = {String-sliding vibrational modes govern the boson peak and phonon anomalies in amorphous materials},
  volume = {25},
  ISSN = {1476-4660},
  url = {http://dx.doi.org/10.1038/s41563-026-02592-9},
  DOI = {10.1038/s41563-026-02592-9},
  number = {7},
  journal = {Nature Materials},
  publisher = {Springer Science and Business Media LLC},
  author = {Xi,  Qing and Wang,  Yinqiao and Tanaka,  Hajime},
  year = {2026},
  
  pages = {1219–1229}
}

@article{Park2026,
  title = {Study of the boson peak in amorphous Ge-Sb-Te using machine learning potentials},
  volume = {114},
  ISSN = {2469-9969},
  url = {http://dx.doi.org/10.1103/2x95-1f6q},
  DOI = {10.1103/2x95-1f6q},
  number = {1},
  journal = {Physical Review B},
  publisher = {American Physical Society (APS)},
  author = {Park,  Pyungjin and Jhi,  Seung-Hoon},
  year = {2026},
  pages = {014201}
}

@article{Wright1993,
  title = {The comparison of molecular dynamics simulations with diffraction experiments},
  volume = {159},
  ISSN = {0022-3093},
  number = {3},
  journal = {Journal of Non-Crystalline Solids},
  publisher = {Elsevier BV},
  author = {Wright,  Adrian C},
  year = {1993},
  month = jan,
  pages = {264–268}
}

@misc{SI,
  title = {},
  note = {See Supplemental Material at [URL will be inserted by the publisher] which contains details about glass preparation procedures and validation of simulated structures, calculation of elastic moduli, Debye-related expressions, calculation of force constants, normal-mode analysis, as well as complementary results including analysis of quasilocalized-vibration core sizes, and scaling of mode characteristics with first phonon-band frequencies.}
}

@article{Baggioli2019,
  title = {Universal Origin of Boson Peak Vibrational Anomalies in Ordered Crystals and in Amorphous Materials},
  volume = {122},
  ISSN = {1079-7114},
  url = {http://dx.doi.org/10.1103/PhysRevLett.122.145501},
  pages={145501},
  DOI = {10.1103/physrevlett.122.145501},
  number = {14},
  journal = {Physical Review Letters},
  publisher = {American Physical Society (APS)},
  author = {Baggioli,  Matteo and Zaccone,  Alessio},
  year = {2019},
  month = Apr 
}

@article{Seyf2017,
  title = {Rethinking phonons: The issue of disorder},
  volume = {3},
  ISSN = {2057-3960},
  url = {http://dx.doi.org/10.1038/s41524-017-0052-9},
  DOI = {10.1038/s41524-017-0052-9},
  number = {1},
  journal = {npj Computational Materials},
  publisher = {Springer Science and Business Media LLC},
  author = {Seyf,  Hamid Reza and Yates,  Luke and Bougher,  Thomas L. and Graham,  Samuel and Cola,  Baratunde A. and Detchprohm,  Theeradetch and Ji,  Mi-Hee and Kim,  Jeomoh and Dupuis,  Russell and Lv,  Wei and Henry,  Asegun},
  year = {2017},
  month = Nov 
}

@article{Huang2021,
  title = {Thermal conductivity modeling on highly disordered crystalline Y1−<i>x</i>Nb<i>x</i>O1.5+<i>x</i>: Beyond the phonon scenario},
  volume = {118},
  ISSN = {1077-3118},
  url = {http://dx.doi.org/10.1063/5.0040546},
  DOI = {10.1063/5.0040546},
  number = {7},
  journal = {Applied Physics Letters},
  pages = {073901},
  publisher = {AIP Publishing},
  author = {Huang,  Muzhang and Liu,  Xiangyang and Zhang,  Peng and Qian,  Xin and Feng,  Yingjie and Li,  Zheng and Pan,  Wei and Wan,  Chunlei},
  year = {2021},
  month = Feb 
}

@article{Zeller1971ThermalSolids,
    title = {{Thermal conductivity and specific heat of non-crystalline solids}},
    year = {1971},
    journal = {Phys. Rev. B},
    author = {Zeller, R. C. and Pohl, R. O.},
    number = {6},
    pages = {2029--2041},
    volume = {4},
    isbn = {0556-2805},
    doi = {10.1016/0375-9601(72)90483-5},
    issn = {03759601},
    pmid = {25246403},
    arxivId = {arXiv:1011.1669v3}
}

@article{baldi2009connection,
  title={Connection between boson peak and elastic properties in silicate glasses},
  author={Baldi, Giacomo and Fontana, Aldo and Monaco, Giulio and Orsingher, Laura and Rols, S and Rossi, Flavio and Ruta, B},
  journal={Physical review letters},
  volume={102},
  number={19},
  pages={195502},
  year={2009},
  publisher={APS}
}

@article{monaco2009anomalous,
  title={Anomalous properties of the acoustic excitations in glasses on the mesoscopic length scale},
  author={Monaco, Giulio and Mossa, Stefano},
  journal={Proceedings of the National Academy of Sciences},
  volume={106},
  number={40},
  pages={16907--16912},
  year={2009},
  publisher={National Academy of Sciences}
}

@article{ruffle2008boson,
  title={Boson peak and its relation to acoustic attenuation in glasses},
  author={Ruffl{\'e}, Beno{\^\i}t and Parshin, DA and Courtens, Eric and Vacher, Ren{\'e}},
  journal={Physical review letters},
  volume={100},
  number={1},
  pages={015501},
  year={2008},
  publisher={APS}
}

@article{aSawamura2018,
  title = {Lateral deformation and defect resistance of compacted silica glass: Quantification of the scratching hardness of brittle glasses},
  volume = {481},
  ISSN = {0022-3093},
  journal = {Journal of Non-Crystalline Solids},
  publisher = {Elsevier BV},
  author = {Sawamura,  Shigeki and Limbach,  René and Behrens,  Harald and Wondraczek,  Lothar},
  year = {2018},
  month = feb,
  pages = {503–511}
}

@article{bndo2018,
  title = {Boson peak,  heterogeneity and intermediate-range order in binary SiO2-Al2O3 glasses},
  volume = {8},
  ISSN = {2045-2322},
  number = {1},
  journal = {Scientific Reports},
  publisher = {Springer Science and Business Media LLC},
  author = {Ando,  Mariana F. and Benzine,  Omar and Pan,  Zhiwen and Garden,  Jean-Luc and Wondraczek,  Katrin and Grimm,  Stephan and Schuster,  Kay and Wondraczek,  Lothar},
  year = {2018},
  month = mar,
  pages = {5394}
}

@article{cdeMacedo2018,
  title = {Lateral hardness and the scratch resistance of glasses in the Na2O-CaO-SiO2 system},
  volume = {492},
  ISSN = {0022-3093},
  journal = {Journal of Non-Crystalline Solids},
  publisher = {Elsevier BV},
  author = {de Macedo,  Guilherme N.B.M. and Sawamura,  Shigeki and Wondraczek,  Lothar},
  year = {2018},
  month = jul,
  pages = {94-101}
}

@article{dLimbach2015,
  title = {Plasticity,  crack initiation and defect resistance in alkali-borosilicate glasses: From normal to anomalous behavior},
  volume = {417–418},
  ISSN = {0022-3093},
  journal = {Journal of Non-Crystalline Solids},
  publisher = {Elsevier BV},
  author = {Limbach,  R. and Winterstein-Beckmann,  A. and Dellith,  J. and M\"{o}ncke,  D. and Wondraczek,  L.},
  year = {2015},
  month = jun,
  pages = {15–27}
}

@article{eSawamura2018,
  title = {Scratch hardness of glass},
  volume = {2},
  ISSN = {2475-9953},
  number = {9},
  journal = {Physical Review Materials},
  publisher = {American Physical Society (APS)},
  author = {Sawamura,  Shigeki and Wondraczek,  Lothar},
  year = {2018},
  month = sep,
  pages = {092601(R)}
}

@article{fPan2021,
  title = {Disorder classification of the vibrational spectra of modern glasses},
  volume = {104},
  ISSN = {2469-9969},
  number = {13},
  journal = {Physical Review B},
  publisher = {American Physical Society (APS)},
  pages = {134106},
  author = {Pan,  Zhiwen and Benzine,  Omar and Sawamura,  Shigeki and Limbach,  Rene and Koike,  Akio and Bennett,  Thomas D. and Wilde,  Gerhard and Schirmacher,  Walter and Wondraczek,  Lothar},
  year = {2021},
  month = oct 
}

@inbook{g1983,
  ISSN = {0921-318X},
  author = {O. V. Mazurin and M. V. Streltsina},
  title = {Handbook of Glass Data-Silica Glass and Binary Silicate Glasses, Part A},
  publisher = {Elsevier},
  year = {1983},
  pages = {326}
}

@article{hMncke2016,
  title = {Transition and post-transition metal ions in borate glasses: Borate ligand speciation,  cluster formation,  and their effect on glass transition and mechanical properties},
  volume = {145},
  ISSN = {1089-7690},
  number = {12},
  pages = {124501},
  journal = {The Journal of Chemical Physics},
  publisher = {AIP Publishing},
  author = {M\"{o}ncke,  D. and Kamitsos,  E. I. and Palles,  D. and Limbach,  R. and Winterstein-Beckmann,  A. and Honma,  T. and Yao,  Z. and Rouxel,  T. and Wondraczek,  L.},
  year = {2016},
  month = sep 
}

@article{refid23Srensen2019,
  title = {Boron anomaly in the thermal conductivity of lithium borate glasses},
  volume = {3},
  ISSN = {2475-9953},
  number = {7},
  journal = {Physical Review Materials},
  publisher = {American Physical Society (APS)},
  author = {Sørensen,  Søren S. and Johra,  Hicham and Mauro,  John C. and Bauchy,  Mathieu and Smedskjaer,  Morten M.},
  year = {2019},
  pages = {075601},
  month = jul 
}

@article{jGriebenow2018,
  title = {Mixed-modifier effect in alkaline earth metaphosphate glasses},
  volume = {481},
  ISSN = {0022-3093},
  journal = {Journal of Non-Crystalline Solids},
  publisher = {Elsevier BV},
  author = {Griebenow,  Kristin and Bragatto,  Caio Barca and Kamitsos,  Efstratios I. and Wondraczek,  Lothar},
  year = {2018},
  month = feb,
  pages = {447–456}
}

\newpage

\onecolumngrid

\includepdf[pages={{},-}]{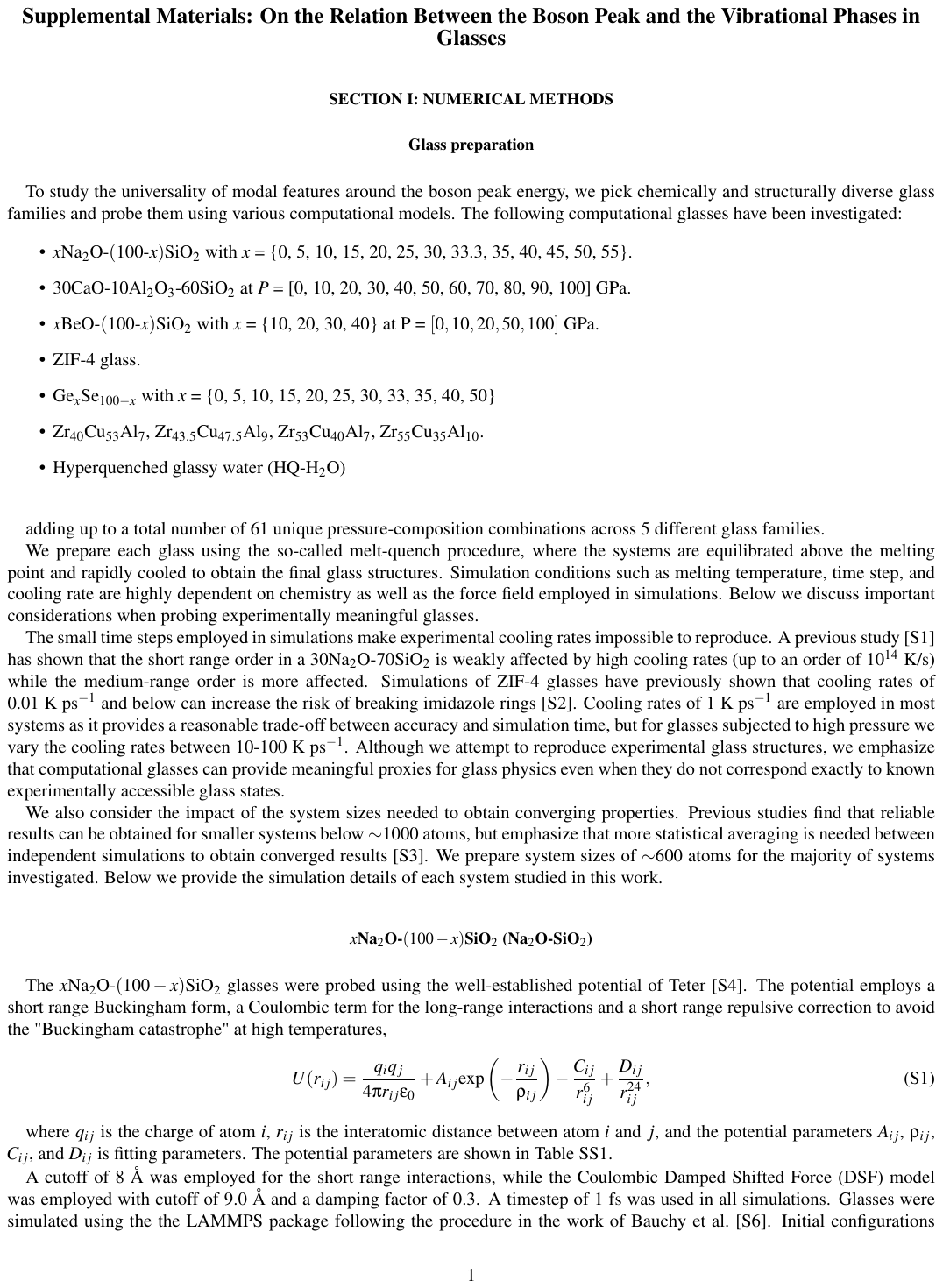}

\end{document}